\documentclass[aps,prl,twocolumn,superscriptaddress,showpacs,preprintnumbers,amsmath,amssymb,floatfix]{revtex4}
\usepackage{graphicx} 
\usepackage{dcolumn}  

\graphicspath{{ps}}

\begin{document}

\preprint{\vbox{ \hbox{   }
                 \hbox{Belle Preprint 2010-19}
		 \hbox{KEK Preprint 2010-30}
                 \hbox{arXiv:0910.4751}
}}

\title{ \quad\\[0.5cm]  First Measurement of Inclusive $B \to X_s \eta$ Decays }

\affiliation{Budker Institute of Nuclear Physics, Novosibirsk}
\affiliation{Faculty of Mathematics and Physics, Charles University, Prague}
\affiliation{Chiba University, Chiba}
\affiliation{University of Cincinnati, Cincinnati, Ohio 45221}
\affiliation{Department of Physics, Fu Jen Catholic University, Taipei}
\affiliation{Justus-Liebig-Universit\"at Gie\ss{}en, Gie\ss{}en}
\affiliation{The Graduate University for Advanced Studies, Hayama}
\affiliation{Hanyang University, Seoul}
\affiliation{University of Hawaii, Honolulu, Hawaii 96822}
\affiliation{High Energy Accelerator Research Organization (KEK), Tsukuba}
\affiliation{Institute of High Energy Physics, Chinese Academy of Sciences, Beijing}
\affiliation{Institute of High Energy Physics, Vienna}
\affiliation{Institute of High Energy Physics, Protvino}
\affiliation{Institute for Theoretical and Experimental Physics, Moscow}
\affiliation{J. Stefan Institute, Ljubljana}
\affiliation{Kanagawa University, Yokohama}
\affiliation{Institut f\"ur Experimentelle Kernphysik, Karlsruher Institut f\"ur Technologie, Karlsruhe}
\affiliation{Korea University, Seoul}
\affiliation{Kyungpook National University, Taegu}
\affiliation{\'Ecole Polytechnique F\'ed\'erale de Lausanne (EPFL), Lausanne}
\affiliation{Faculty of Mathematics and Physics, University of Ljubljana, Ljubljana}
\affiliation{University of Maribor, Maribor}
\affiliation{Max-Planck-Institut f\"ur Physik, M\"unchen}
\affiliation{University of Melbourne, School of Physics, Victoria 3010}
\affiliation{Nagoya University, Nagoya}
\affiliation{Nara Women's University, Nara}
\affiliation{National Central University, Chung-li}
\affiliation{National United University, Miao Li}
\affiliation{Department of Physics, National Taiwan University, Taipei}
\affiliation{H. Niewodniczanski Institute of Nuclear Physics, Krakow}
\affiliation{Nippon Dental University, Niigata}
\affiliation{Niigata University, Niigata}
\affiliation{Novosibirsk State University, Novosibirsk}
\affiliation{Osaka City University, Osaka}
\affiliation{Panjab University, Chandigarh}
\affiliation{Saga University, Saga}
\affiliation{University of Science and Technology of China, Hefei}
\affiliation{Seoul National University, Seoul}
\affiliation{Sungkyunkwan University, Suwon}
\affiliation{School of Physics, University of Sydney, NSW 2006}
\affiliation{Tata Institute of Fundamental Research, Mumbai}
\affiliation{Excellence Cluster Universe, Technische Universit\"at M\"unchen, Garching}
\affiliation{Toho University, Funabashi}
\affiliation{Tohoku Gakuin University, Tagajo}
\affiliation{Tohoku University, Sendai}
\affiliation{Department of Physics, University of Tokyo, Tokyo}
\affiliation{Tokyo Metropolitan University, Tokyo}
\affiliation{IPNAS, Virginia Polytechnic Institute and State University, Blacksburg, Virginia 24061}
\affiliation{Wayne State University, Detroit, Michigan 48202}
\affiliation{Yonsei University, Seoul}
  \author{K.~Nishimura}\affiliation{University of Hawaii, Honolulu, Hawaii 96822} 
  \author{T.~E.~Browder}\affiliation{University of Hawaii, Honolulu, Hawaii 96822} 
  \author{I.~Adachi}\affiliation{High Energy Accelerator Research Organization (KEK), Tsukuba} 
  \author{H.~Aihara}\affiliation{Department of Physics, University of Tokyo, Tokyo} 
 \author{K.~Arinstein}\affiliation{Budker Institute of Nuclear Physics, Novosibirsk}\affiliation{Novosibirsk State University, Novosibirsk} 
  \author{T.~Aushev}\affiliation{\'Ecole Polytechnique F\'ed\'erale de Lausanne (EPFL), Lausanne}\affiliation{Institute for Theoretical and Experimental Physics, Moscow} 
  \author{A.~M.~Bakich}\affiliation{School of Physics, University of Sydney, NSW 2006} 
  \author{V.~Balagura}\affiliation{Institute for Theoretical and Experimental Physics, Moscow} 
  \author{E.~Barberio}\affiliation{University of Melbourne, School of Physics, Victoria 3010} 
  \author{K.~Belous}\affiliation{Institute of High Energy Physics, Protvino} 
% ...
\collaboration{The Belle Collaboration}

\include{pub327}

\noaffiliation

\begin{abstract}
We report a first measurement of inclusive $B \to X_s \eta$ decays, where 
$X_s$ is a charmless state with unit strangeness. The measurement is based on  
a pseudo-inclusive reconstruction technique and uses a sample of 
$657 \times 10^6~B\bar{B}$ pairs accumulated with the Belle 
detector at the KEKB $e^+e^-$ collider.  For $M_{X_s} < 2.6 ~\mathrm{GeV}/c^2$, 
we measure a branching fraction of 
$(26.1 \pm 3.0 (\mathrm{stat}) ^{+1.9}_{-2.1} (\mathrm{syst}) ^{+4.0}_{-7.1} (\mathrm{model})) \times 10^{-5}$ and a direct $CP$ asymmetry of
$\mathcal{A}_{CP} = -0.13 \pm 0.04 ^{+0.02}_{-0.03}$. Over half
of the signal occurs in the range $M_{X_s} > 1.8 ~\mathrm{GeV}/c^2$.
\end{abstract}


\maketitle

{\renewcommand{\thefootnote}{\fnsymbol{footnote}}}
\setcounter{footnote}{0}

Decays of $B$ mesons involving the $b \to s$ transition are an excellent
tool for searches for physics beyond the Standard Model (SM).
Theoretical treatments of these decays into exclusive hadronic final states, 
however, suffer from large uncertainties in the hadronization process. The 
uncertainties can be 
effectively reduced by leaving some of the final states in the calculation at 
the quark level, which corresponds to a measurement of an inclusive hadronic 
state $X_s$ of unit strangeness.

Among such $b \to s$ decays, 
those involving the $\eta$ and $\eta^\prime$ mesons exhibit
unique properties due to interference between their underlying SU(3)
octet and singlet components \cite{lipkin}. The CLEO collaboration reported the
first measurement of inclusive $B \to X_s \eta^\prime$ with an unexpectedly 
large branching fraction and an $X_s$ spectrum that peaks at high $X_s$ mass 
\cite{CLEO_first_EtaPrimeXs}, a result confirmed in improved, higher-statistics 
measurements \cite{CLEO_second_EtaPrimeXs,BaBar_EtaPrimeXs}. 
Explanations included a large 
intrinsic $c\bar{c}$ component of the $\eta^\prime$ \cite{intrinsic_charm}, 
the QCD anomaly mechanism \cite{atwood_soni_QCD_anomaly} that couples two 
gluons to the flavor singlet component of the $\eta^\prime$, and 
also new physics sources \cite{Hou_NP}. The first is disfavored by the lack of  
an enhancement of $B \to \eta_c K$ relative to $B \to J/\psi K$ 
\cite{CLEO_JPsi_K}, while the second is disfavored by a measurement of 
$\Upsilon (1S) \to \eta^\prime X$ \cite{Y1S_etagg}, which indicates an 
$\eta^\prime g g$ form factor that cannot explain the enhancement. A recent 
treatment \cite{charming_penguin} using soft collinear effective theory  
suggests that a measurement of the complementary process $B \to X_s \eta$ can 
elucidate the possible contribution from nonperturbative charm-penguin 
amplitudes or 
higher-order gluonic operators to both the $\eta$ and $\eta^\prime$ processes.  
CLEO performed the only previous search with an upper limit of 
$\mathcal{B}(B \to X_s \eta) <4.4 \times 10^{-4}$ \cite{CLEO_first_EtaPrimeXs}.

In this Letter, we report a measurement of $B \to X_s \eta$ using a 
sample of $657 \times 10^6~B\bar{B}$ pairs accumulated with the Belle 
detector at the KEKB $e^+e^-$ collider \cite{KEKB}.  The Belle detector is a 
large solid-angle magnetic spectrometer and is described in detail 
elsewhere \cite{Belle}.

We reconstruct candidate $B$ mesons using a pseudo-inclusive method, with the 
$X_s$ composed of a $K^+$ or $K_S^0 (\to \pi^+\pi^-)$ and up to four 
pions, of which at most one is a $\pi^0 (\to \gamma\gamma)$.  This gives a 
total of 18 reconstructed channels and their charge-conjugates \cite{CC}. 
Candidate $\eta$ mesons are reconstructed in the $\eta \to \gamma\gamma$ mode 
from photons with $E_\gamma > 200~\mathrm{MeV}$.  The invariant mass of the 
$\gamma$-pair is required to lie between $520 ~\mathrm{MeV/}c^2$ and 
$570 ~\mathrm{MeV/}c^2$, or within 2$\sigma$ of the nominal mass.  We veto an 
$\eta$ candidate if either of its photons 
can be combined with another photon in the event to form a candidate $\pi^0$.  
To suppress background from radiative $B$ decays, we require the 
energy asymmetry of the two photons, defined as 
$|E_{\gamma_1} - E_{\gamma_2}| / |E_{\gamma_1} + E_{\gamma_2}|$, to be less than 0.6.  
The momenta of $\eta$ candidates are recalculated using 
the nominal $\eta$ mass \cite{PDG}.  To suppress secondary $\eta$-mesons from 
$b \to c \to \eta$ chains, we retain only $\eta$ candidates whose 
center-of-mass (CM) momentum satisfies 
$|{\bf p}_{\eta}^*| > 2.0 ~\mathrm{GeV}/c$.

Charged pions and kaons are selected based on  
information from the time-of-flight, aerogel Cherenkov, and drift chamber 
$dE/dx$ systems.  Typical efficiencies to correctly identify kaons (pions)  
are above 88\% (98\%), with misidentification rates for pions 
as kaons (kaons as pions) below 12\% (4\%). 
$K_S^0$ candidates are required to have an invariant mass within 16 MeV/$c^2$ 
(4$\sigma$) of the $K_S^0$ mass and a displaced vertex from the interaction 
point.   
For $\pi^0$ candidates, each daughter photon is required to have energy 
greater than $50~(100) ~\mathrm{MeV}$ in the barrel (endcap) region and 
a shower shape consistent with a photon.  The invariant mass of the photon 
pair must be within 15 MeV/$c^2$ (2.5$\sigma$) of the $\pi^0$ mass.  The 
$\pi^0$ momentum is recalculated using the nominal $\pi^0$ mass.  To suppress 
combinatorial backgrounds, we require $\pi^0$ candidates to have laboratory 
momenta greater than $300 ~\mathrm{MeV}/c$.

Pions and kaons are combined to form an $X_s$, and $B$ meson candidates are  
formed from combinations of an $\eta$ and an $X_s$.  A beam-constrained mass, 
$M_{\mathrm{bc}} = \sqrt{E^2_{\mathrm{beam}}/c^4 - |{\bf {p}}_B^*|^2/c^2}$ and 
energy difference, $\Delta{E} = E_B - E_{\mathrm{beam}}$ are calculated, where 
$E_{\mathrm{beam}}$, ${\bf {p}}_B^*$, and $E_{B}$ are the beam energy, $B$ 
momentum, and $B$ energy, all in the CM frame.  The signal is obtained using 
fits to $M_{\mathrm{bc}}$ with $|\Delta{E}| < 0.1~\mathrm{GeV}$.

We use a simulated signal Monte Carlo (MC) sample \cite{MC_GEN} consisting of 
$B \to K \eta$ for $M_{X_s} < 0.6 ~\mathrm{GeV}/c^2$, $B \to K^* \eta$ for 
$M_{X_s} \in [0.8,1.0] ~\mathrm{GeV}/c^2$, and $B \to X_s \eta$ 
in all other mass regions ($M_{X_s} \in [0.6,0.8] ~\mathrm{GeV}/c^2$, 
and $M_{X_s} > 1.0 ~\mathrm{GeV}/c^2$).  For the $B \to X_s \eta$ component, 
fragmentation of the $X_s$ system into hadrons is simulated by 
PYTHIA \cite{PYTHIA}, assuming a model in which the $X_s$ mass spectrum 
is flat from the $K\pi$ threshold up to $3.2 ~\mathrm{GeV}/c^2$.   
We find an average of approximately nine $B$ candidates per event, of which 
we select the candidate with the lowest $\chi^2$.  This $\chi^2$ is defined as 
the sum of $\chi^2_{\Delta{E}} = (\Delta{E}/\sigma_{\Delta{E}})^2$, 
where the resolution $\sigma_{\Delta_E}$ is estimated separately for each 
reconstructed mode, 
and, if available, a reduced-$\chi^2$ of a vertex fit that includes all $X_s$ 
daughter charged tracks except those used as part of a $K_S^0$ candidate.  
After applying this procedure, we select the correctly reconstructed 
$B$ in 56\% of simulated events.

The dominant background to $B \to X_s \eta$ comes from continuum production of 
quark pairs, $e^+e^- \to q\bar{q}~(q = u,d,s,c)$.  These events have a 
jet-like topology, and are suppressed relative to the spherical 
$B\bar{B}$ events using a Fisher discriminant \cite{Fisher} formed from 
event shape variables \cite{SFW,KSFW}.  Further suppression is obtained by 
combining this Fisher discriminant with the cosine of the $B$ flight 
direction in the CM frame and, when available, the displacement between the 
signal $B$ and the other $B$ in the event. This suppression is optimized as a 
function of $b$-flavor tag quality \cite{TaggingNIM}, and is approximately 
34\% efficient for the signal modes while suppressing over 99\% of the 
continuum background.

Decays of the type $B \to X_c \eta$ and $B \to X_c \to X_s \eta$, where $X_c$ 
is any state containing charm mesons, may have final states 
identical to the signal mode.  We search among the candidate $B$ decay 
products for combinations consistent with selected charm meson decays and  
veto the candidate if the mass of the reconstructed combination is within 
$\pm 2.5 \sigma$ of the known mass.  
The modes and their veto widths are: 
$D^0 \to K n\pi^\pm (\pi^0)$, 13.5 (44.5) MeV$/c^2$; 
$D^+ \to K n\pi^\pm (\pi^0)$, 12.5 (31.3) MeV$/c^2$;
$D^0 \to K_S^0 \eta $, 31.3 MeV$/c^2$;
$D_s^+ \to \eta \pi^+ $, 29.3 MeV$/c^2$; and
$\eta_{c}(1S) \to \eta \pi^+ \pi^- $, 85.0 MeV$/c^2$.
We also veto events with an $\eta^\prime \to \eta\pi^+\pi^-$ candidate with 
an invariant mass $M_{\eta\pi\pi}$ within 100 MeV$/c^2$ of the nominal 
$\eta^\prime$ mass.

Signal yields are obtained using an extended unbinned maximum likelihood fit 
to $M_{\mathrm{bc}}$ in $200 ~\mathrm{MeV}/c^2$ bins of 
$X_s$ mass up to $2.6 ~\mathrm{GeV}/c^2$.  The probability 
density function (PDF) for the signal is taken as a Gaussian, with the 
mean and width determined from the appropriate signal MC sample 
($K\eta$, $K^*\eta$, or $X_s\eta$) for the mass bin, with calibration factors 
taken from a $B \to D \pi$ control sample.  All reconstructed modes 
are combined for the fit, and no attempt is made to separate correctly 
reconstructed $B$ candidates and those with some missing or incorrectly 
attributed $B$ daughters (self-cross-feed). Shapes for the charm contributions 
remaining after the vetoes are assigned based on a MC sample of generic 
$b \to c$ processes.  Four separate PDFs are assigned for the largest charm 
backgrounds as identified in MC: 
$B^0 \to \bar{D}{}^0 \eta$, $B^0 \to \bar{D}{}^{*0}\eta$, 
$B^0 \to {D}^{(*)-} \pi^+ \eta$, and $B^+ \to \bar{D}{}^{(*)0} \pi^+ \eta$.  
All other $b \to c$ backgrounds are combined into another PDF.  Each charm PDF 
consists of a Gaussian component to describe the peaking in $M_{\mathrm{bc}}$, 
and an empirically determined 
parameterization (ARGUS function) \cite{ARGUS} to describe non-peaking 
combinatorial contributions.  The shape parameters are taken from the 
appropriate background MC sample.  Normalizations of the modes 
$B^0 \to \bar{D}{}^{(*)0} \eta$ are based on the previous Belle 
measurement \cite{Blyth}.  The branching fractions for the decays 
$B \to D^{(*)} \pi \eta$ are unknown, so their normalization is 
determined by a simultaneous $\chi^2$ minimization based on the difference 
between the expected and observed $M_{\mathrm{bc}}$ distribution of the events 
in all eight veto windows.  The relative normalizations of 
$D^{-}\pi^+\eta$ and $D^{*-}\pi^+\eta$ are assumed to be the same, and likewise 
for the $D^{(*)0}$ modes.  The $\chi^2$ technique is verified by repeating 
the optimization over the $B^0 \to \bar{D}{}^{(*)0} \eta$ modes, for which the 
results are consistent with the previous Belle measurement. This $\chi^2$ is 
also used to study systematic errors on the normalizations of all charm PDFs. 
Normalization for the PDF that includes all other 
$b \to c$ modes is fixed to the MC expectation.  The remaining combinatorial 
$q\bar{q}$ backgrounds are modeled with an ARGUS function.  For the final 
fit, the signal yield and both the yield and shape parameter of the 
$q\bar{q}$ ARGUS PDF are allowed to vary.

Rare $B$ decay backgrounds are studied with a dedicated MC 
sample, and include contributions from $B \to X_s \eta^\prime$, 
$B \to X_s \gamma$, and $B \to X_d \eta$.  These expected yields are 
subtracted from the fit yield to give a final yield.  The expected yields for 
$B \to X_s \eta^\prime$ and $B \to X_s \gamma$ are based on the known branching 
fractions, and are found to be less than 0.5 events in 
each $X_s$ mass bin.  The $B^+ \to \pi^+ \eta$ branching fraction is also 
known, and the expectation is 5.2 events in the lowest bin of $X_s$ 
mass.  We estimate the contribution from other $B \to X_d \eta$ modes by 
repeating the reconstruction
and the fitting procedure but replacing the $K^+$ candidate of $X_s$ with a 
$\pi^+$ candidate.  
Performing these fits on data and using a dedicated $X_d \eta$ MC sample to 
estimate the rate to misreconstruct $X_d$ as $X_s$, we estimate a total 
contamination of $19.1 \pm 2.3$ events from $X_d \eta$, distributed uniformly 
in the range $M_{X_s} \in [0.6,2.6] ~\mathrm{GeV}/c^2$.

\begin{figure}[htb]
\begin{center}
  \begin{tabular}{cc}
    \hspace{-2mm}\includegraphics[width=0.26\textwidth]{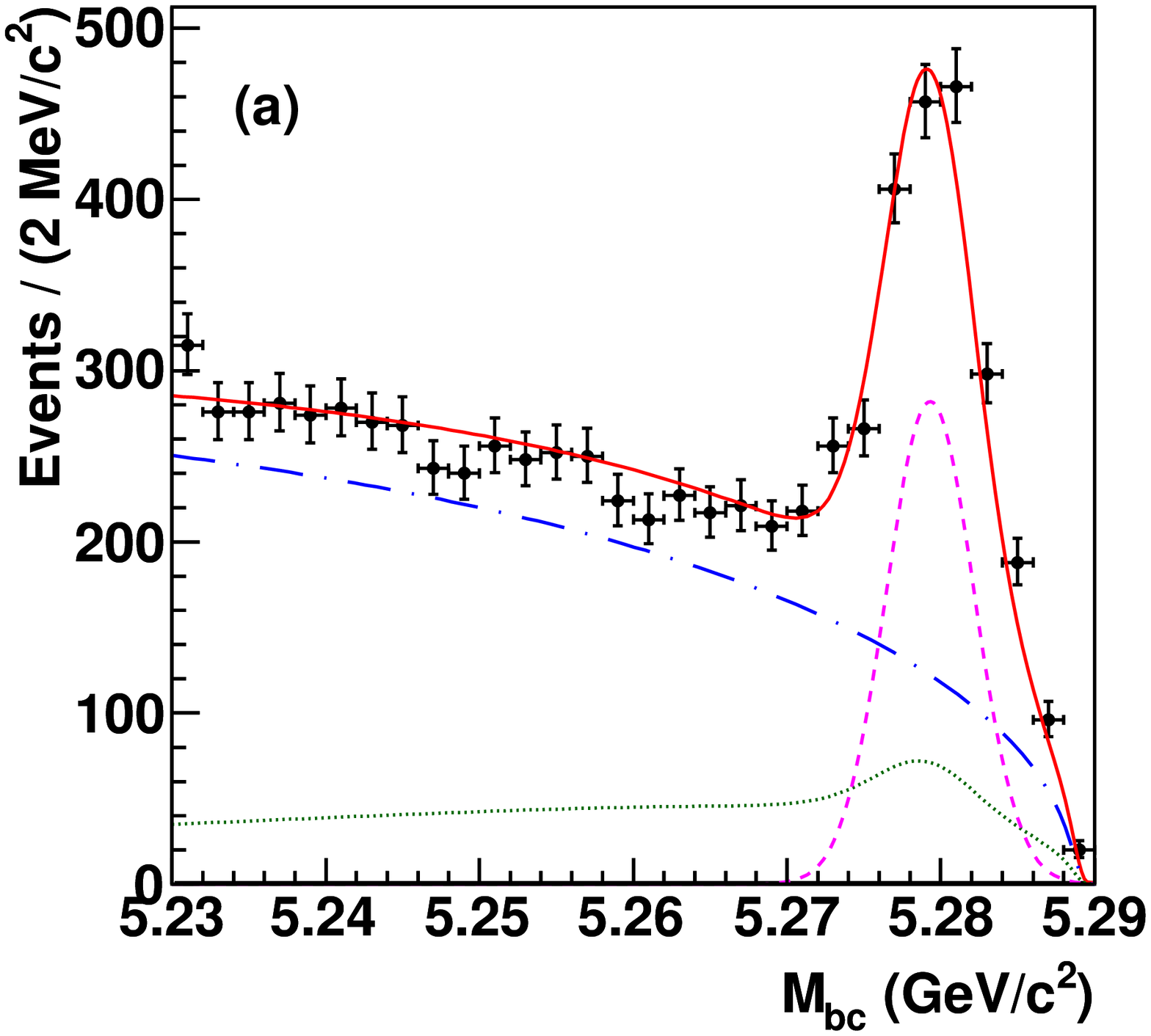} & 
    \hspace{-4mm}\includegraphics[width=0.26\textwidth]{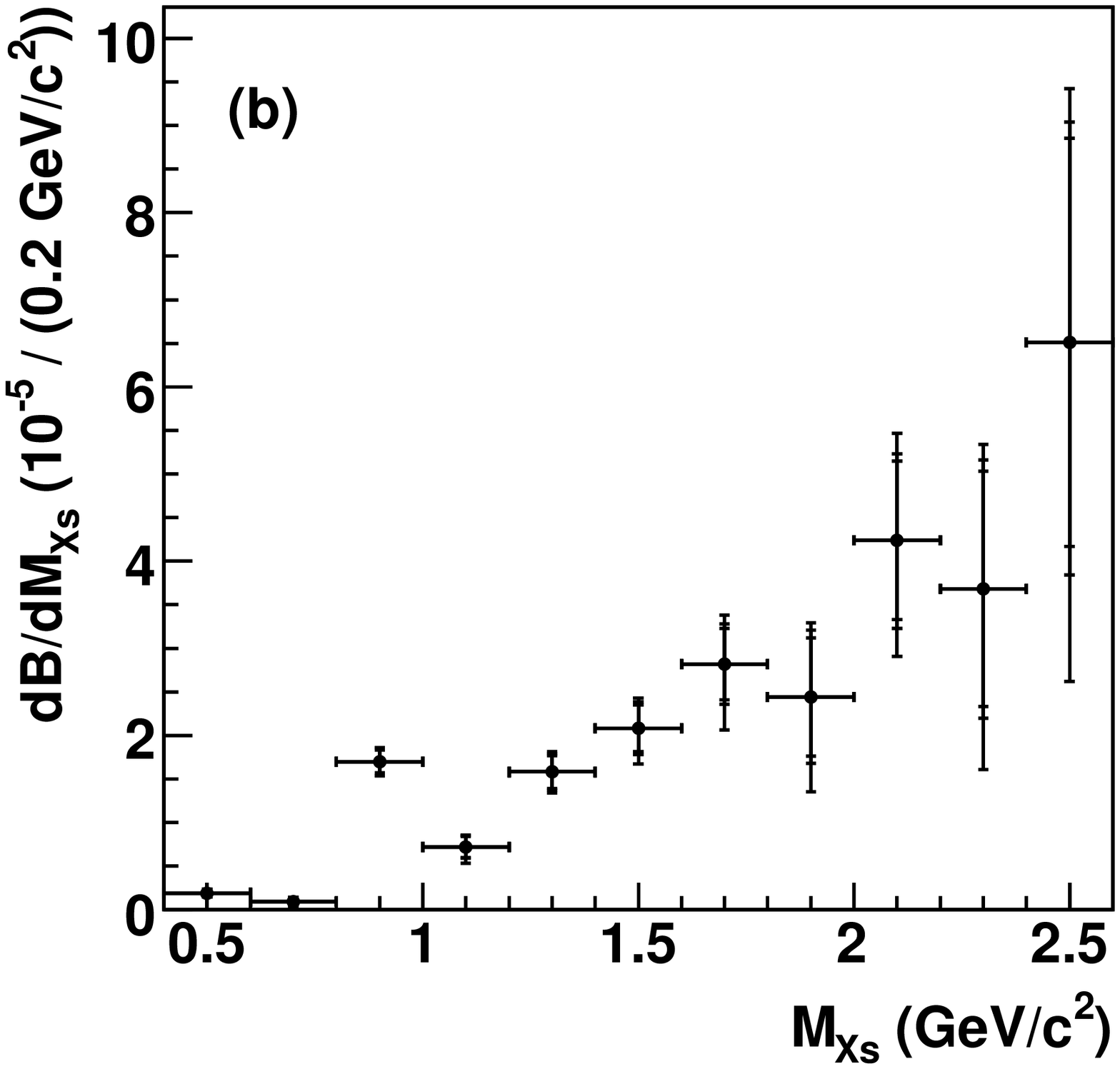} \\
  \end{tabular}
\end{center}
\caption{ (color online). 
(a) The $M_{\mathrm{bc}}$ distribution for the full mass range, 
$M_{X_s} \in [0.4,2.6] ~\mathrm{GeV}/c^2$.  The points with 
errors correspond to the data, while the curves correspond to the overall fit 
PDF (solid red), the signal PDF (dashed magenta), the sum of all $b \to c$ 
background PDFs (dotted green), and the combinatorial background PDF
(dash-dotted blue). (b) Differential branching fraction, 
$d\mathcal{B}/dM_{X_s}$, for $B \to X_s \eta$.  The three error bars correspond 
to statistical error only, statistical plus systematic error, and total error.}
\label{fig1}
\end{figure}

The fit to the full mass range, $M_{X_s} \in [0.4,2.6]~\mathrm{GeV}/c^2$, 
is shown in Fig.~\ref{fig1}(a), and gives a background-subtracted yield 
 of $1054 \pm 54 ^{+16}_{-18}$.  We also define a high mass region, 
$M_{X_s} \in [1.8,2.6] ~\mathrm{GeV}/c^2$, where the summed yield 
is  $233 \pm 34 ^{+13}_{-15}$.  Significances are determined in each mass 
bin by convolving the likelihood function with a Gaussian of width determined 
by the systematic errors on the yield.  The maximum likelihood, 
$\mathcal{L}_{\mathrm{max}}$, and the likelihood at a signal yield of zero, 
$\mathcal{L}_{0}$, are used to determine the significance, which is defined as  
$\sqrt{-2\ln(\mathcal{L}_0/\mathcal{L}_{\mathrm{max}})}$. The 
significance is 23 (7) for the full (high) $X_s$ mass range.

Reconstruction efficiencies in bins of $X_s$ mass range from 6.5\% to 0.1\%, 
not including the branching fraction for $\eta \to \gamma\gamma$; these results 
are based on the signal MC and assume equal production of $B^+B^-$ and 
$B^0\bar{B^0}$ at $\Upsilon (4S)$.  Figure~\ref{fig1}(b) shows the differential 
branching fraction as a function of $M_{X_s}$.  Table \ref{tab:Summary_Table} 
gives the final results for each $X_s$ mass bin.  For the 
full $M_{X_s}$ range, we sum the individual contributions and find the 
following partial branching fraction
$\mathcal{B}(B \to X_s \eta; M_{X_s} \in [0.4,2.6] ~\mathrm{GeV}/c^2) = (26.1 \pm 3.0 {^{+1.9}_{-2.1}} {^{+4.0}_{-7.1}}) \times 10^{-5}$, where errors are statistical,
(model-independent) systematic, and decay modeling.  
A large fraction of the inclusive signal occurs in the high 
mass region, where we find 
$\mathcal{B}(B \to X_s \eta; M_{X_s} \in [1.8,2.6] ~\mathrm{GeV}/c^2) = (16.9 \pm 2.9~\mathrm{(stat)} ^{+1.5}_{-1.8}~\mathrm{(syst)} ^{+3.3}_{-5.9}~\mathrm{(model)}) \times 10^{-5}$.

The direct $CP$ asymmetry is defined as
$\mathcal{A}_{CP} = (\mathcal{B}^- - \mathcal{B}^+)/(\mathcal{B}^- + \mathcal{B}^+)$, where $\mathcal{B}^+ (\mathcal{B}^-)$  is the partial branching 
fraction for $B^+$ or $B^0$ ($B^-$ or $\bar{B^0}$).
We measure this asymmetry in the 
subset of reconstructed modes in which the $B$ flavor can be inferred from the 
final state.  We correct the fitted $CP$ asymmetry to account for events that 
are reconstructed with the wrong $B$ flavor by multiplying the raw fitted 
asymmetry by a dilution factor.  This factor is estimated from 
the signal MC, and ranges from unity to 1.05. The bin-by-bin results, as well 
as the results of separate fits for $\mathcal{A}_{CP}$ over the full 
$X_s$ mass range and the range above the narrow kaonic resonances 
($M_{X_s} \in [1.0,2.6] ~\mathrm{GeV}/c^2$), are shown in 
Table \ref{tab:Summary_Table}.  For $M_{X_s} \in [0.4,2.6]~\mathrm{GeV}/c^2$, 
we find $\mathcal{A}_{CP} = -0.13 \pm 0.04 ^{+0.02}_{-0.03}$, with a significance 
of 2.6$\sigma$ relative to a null asymmetry. 
All $\mathcal{A}_{CP}$ results that include the range 
$M_{X_s} \in [0.4,0.6] ~\mathrm{GeV}/c^2$ are calculated with 
the assumption that the $B^+ \to \pi^+ \eta$ backgrounds in this region  
contribute a $CP$ asymmetry consistent with the existing measured world 
average \cite{PDG}.

\begin{table}[t]
\caption{Measured background-subtracted signal yields ($N_S$), 
branching fractions ($\mathcal{B}$), 
and $CP$ asymmetry ($\mathcal{A}_{CP}$), for each $M_{X_s}$ range.
Uncertainties on $N_S$ are statistical.  Uncertainties on $\mathcal{B}$ are 
statistical, systematic, and modeling, respectively. 
The uncertainties for $\mathcal{A}_{CP}$ are statistical and systematic.}
\label{tab:Summary_Table}
\begin{tabular}
{c c c c c}
\hline \hline
$M_{X_s} (\mathrm{GeV}/c^2)$ & 
$N_S$ & 
$\mathcal{B} (10^{-6})$ & 
$\mathcal{A}_{CP} (10^{-2})$\\

\hline
0.4--0.6 & $60   \pm 12$ & 
           $1.9  \pm 0.4  \pm 0.1         \pm 0.0$ &
           $-35  \pm 18   \pm 2$\\

0.6--0.8 & $15   \pm  9$ & 
           $0.9  \pm 0.5  \pm 0.1         ^{+0.1}_{-0.0}$ &
           $2    \pm  40  \pm 13$\\

0.8--1.0 & $250  \pm 19$ & 
           $17.0 \pm 1.3  ^{+0.9}_{-1.0}     \pm 0.0$ &
           $-4   \pm 7    \pm 2$\\

1.0--1.2 & $84   \pm 14$ &
           $7.2  \pm 1.2  {^{+0.4}_{-0.5}}   {^{+0.3}_{-1.4}}$ &
           $-26  \pm 15   ^{+3}_{-4}$\\

1.2--1.4 & $146  \pm 17$ & 
           $15.8 \pm 1.9  \pm 1.0         ^{+1.0}_{-1.1}$ &
           $-22  \pm 11   ^{+2}_{-3}$\\

1.4--1.6 & $137  \pm 18$ &
           $20.8 \pm 2.7  {^{+1.3}_{-1.4}}   {^{+1.9}_{-2.8}}$ & 
           $-15  \pm 12   ^{+2}_{-3}$\\

1.6--1.8 & $128  \pm 18$ & 
           $28.2 \pm 4.1  \pm 2.1        ^{+3.3}_{-6.1}$ & 
           $-25  \pm 13   ^{+2}_{-3}$\\

1.8--2.0 & $64   \pm 18$ &
           $24.4 \pm 6.8  {^{+3.6}_{-3.4}}  {^{+3.7}_{-7.8}}$ & 
           $-31  \pm 26   \pm 6$\\

2.0--2.2 & $86   \pm 18$ & 
           $42.4 \pm 9.1  {^{+3.8}_{-4.3}}  {^{+7.3}_{-8.7}}$ & 
           $34   \pm 20   ^{+4}_{-3}$\\

2.2--2.4 & $49   \pm 18$ & 
           $36.8 \pm 13.5 {^{+5.9}_{-6.1}}  {^{+7.6}_{-14.5}}$ & 
           $2    \pm 32   \pm 5$\\

2.4--2.6 & $35   \pm 13$ & 
           $65.1 \pm 23.4 {^{+9.5}_{-12.8}} {^{+14.5}_{-28.3}}$ & 
           $-40  \pm 36   ^{+7}_{-12}$\\
\hline
0.4--2.6 & $1053 \pm 54$ &
           $261 \pm 30 {^{+19}_{-21}} {^{+40}_{-71}}$ &
           $-13  \pm 4 ^{+2}_{-3}$ \\
1.0--2.6 & $728  \pm 48$ &
           $241 \pm 30 {^{+18}_{-20}} {^{+40}_{-71}}$ &
           $-15  \pm 6 \pm 3$\\
1.8--2.6 & $233  \pm 34$ &
           $169 \pm 29 {^{+15}_{-18}} {^{+33}_{-59}}$ &
           $0   \pm 14 \pm 5$\\
\hline \hline
\end{tabular}
\end{table}

Systematic errors on the fitted signal yield are dominated by 
PDF uncertainties. Uncertainties in the signal PDF parameters are studied 
using a $B \to D \pi$ control sample, while those due to normalizations and 
shapes for the $b \to c$ backgrounds are estimated using comparisons between 
the veto window $\chi^2$ procedure and either MC expectations or, 
when available, previous measurements.  Errors
from the background subtractions are dominated by uncertainties in the 
estimate of backgrounds from $B \to X_d \eta$.  Our estimates of these 
backgrounds may have included other small contributions, such as those from 
$B \to X_s \eta$, so we allow these 
estimates to vary by $-100\%$. Positive uncertainties are estimated from the 
difference in expected yields assuming a flat distribution of $X_d$ events 
in $X_s$ mass versus those obtained from a MC study of cross-feed from $X_d$ 
mass to $X_s$ mass.  
In all cases the systematic uncertainties on the background-subtracted signal 
yields are at least a factor of two smaller than the statistical errors.

The model-independent systematic error includes contributions from the signal 
yield, the selection efficiency, the number of $B\bar{B}$ pairs, and the 
$\eta \to \gamma\gamma$ branching fraction \cite{PDG}.  For $X_s$ mass bins 
above $1.8~\mathrm{GeV}/c^2$, the errors on the signal yields from 
uncertainties 
in the PDF shapes (primarily for the charm PDFs) dominate with a contributed 
relative uncertainty of 7-18\%.  
For the lower $X_s$ mass bins, the efficiency error is the largest 
contribution with a relative uncertainty of 5-6\%.  This error is the 
combination of 
individually determined contributions from control sample studies of the 
following: tracking, reconstruction of $\pi^0$, $\eta$, and $K_S^0$, 
particle identification, continuum suppression, and candidate selection.  

We define an additional error due to modeling of the $X_s$ system, which is 
studied in three parts.  The first is due to the fraction of unreconstructed 
modes (e.g., modes with too many total or neutral pions or additional kaons). 
We vary these fractions by $\pm30\%$ of the PYTHIA expectation and use the 
differences in efficiency to estimate an $M_{X_s}$ bin-dependent uncertainty 
that rises with $X_s$ mass from zero to $\pm21.1\%$. 
The second is due to differences in the observed frequency of decay modes and 
those expected from PYTHIA.  We find good agreement between data and MC in the 
relative amounts of charged and neutral $B$ modes, modes with $K_S^0$ and 
those with $K^+$, and modes with one or two total $\pi$'s and those with three 
or four total $\pi$'s.  However, we find a significant excess of modes 
without a $\pi^0$ over those with a $\pi^0$, which we attribute to inaccuracies 
in the PYTHIA fragmentation.  To quantify this uncertainty, we re-estimate 
the PYTHIA efficiencies with the fraction of $\pi^0$ modes adjusted to match 
data, and use the difference between this value and the nominal efficiency 
to assign an error.  This error is usually only negative, due to the higher 
reconstruction efficiency for modes without a $\pi^0$, and is as large as 
$-37\%$ in the highest $X_s$ mass bin.  The final component of the modeling 
uncertainty is due to the assumed $X_s$ mass spectrum.  We study the 
efficiencies of other $M_{X_s}$ signal MC samples where the spectrum rises 
toward high mass and assign errors based on the differences from the 
flat $M_{X_s}$ MC. Using these samples, we also study the fractions of  
self-cross-feed candidates that are reconstructed with an incorrect  
$X_s$ mass. These effects are small compared to the first two components of  
the modeling error.

The systematic error on $\mathcal{A}_{CP}$ includes contributions due to: 
uncertainties in the PDF parameters; possible detector 
and measurement biases, which are estimated 
from the measured $\mathcal{A}_{CP}$ of the $B \to D \pi$ control 
sample and the signal MC, respectively; uncertainty due to the signal model is 
studied by checking the 
fractions of events with incorrectly identified flavor using alternative 
$M_{X_s}$ spectra; and possible contamination due to 
$B \to \pi \eta~(B \to X_d \eta)$ decays is estimated by varying their 
$A_{CP}$ by the measured uncertainty \cite{PDG} ($\pm100\%$).
 
In summary, we report the first measurement of the inclusive process 
$B \to X_s \eta$, and find a partial branching fraction of 
$\mathcal{B}(B \to X_s \eta; M_{X_s} \in [0.4,2.6] ~\mathrm{GeV}/c^2) = (26.1 \pm 3.0 (\mathrm{stat}) ^{+1.9}_{-2.1} (\mathrm{syst}) ^{+4.0}_{-7.1} (\mathrm{model})) \times 10^{-5}$.  
The measured $M_{X_s}$ dependent branching fractions are consistent with 
the known $B \to K \eta$ and $B \to K^*(892) \eta$ processes \cite{CLEO_Keta}. 
 In the high mass region, $M_{X_s} \in [1.8,2.6] ~\mathrm{GeV}/c^2$, which is
 above any significant contributions from previously measured exclusive 
processes \cite{BaBar_Kst1430}, we observe a signal with a
 $7 \sigma$ significance.  We also measure 
the $CP$ asymmetry of $B \to X_s \eta$, both as a function of $M_{X_s}$ and for 
the full mass range, where we find 
$\mathcal{A}_{CP} = -0.13 \pm 0.04 ^{+0.02}_{-0.03}$.
No theoretical prediction is currently available for the shape of the $M_{X_s}$ 
spectrum.  However, the similarity in spectral shape to 
$B \to X_s \eta^\prime$ and the lack of strong suppression of the 
$B \to X_s \eta$ branching fraction relative to the $\eta^\prime$ mode imply 
that the origin of the large contribution in the $\eta^\prime$ mode is also 
common to the $\eta$ mode \cite{charming_penguin}, and disfavors 
$\eta^\prime$ specific mechanisms \cite{intrinsic_charm,atwood_soni_QCD_anomaly}.

\vspace{0.3cm}
We thank the KEKB group for excellent operation of the
accelerator, the KEK cryogenics group for efficient solenoid
operations, and the KEK computer group and
the NII for valuable computing and SINET3 network support.  
We acknowledge support from MEXT, JSPS and Nagoya's TLPRC (Japan);
ARC and DIISR (Australia); NSFC (China); MSMT (Czechia);
DST (India); MEST, NRF, NSDC of KISTI (Korea); MNiSW (Poland); 
MES and RFAAE (Russia); ARRS (Slovenia); SNSF (Switzerland); 
NSC and MOE (Taiwan); and DOE (USA).




\begin{thebibliography}{99}
\bibitem{CC}
Throughout this paper, 
the inclusion of the charge-conjugate mode decay is implied
unless otherwise stated.
\end{thebibliography}
\end{document}